\documentclass[10pt,letterpaper,twocolumn]{article} 

\usepackage{ol2}
\usepackage[draft]{hyperref}
\usepackage{amsmath}
\usepackage{graphicx,bm}
\usepackage[normalem]{ulem}

\begin{document}

\twocolumn[ 

\title{Limits to superweak amplification of beam shifts}

\author{J\"org B G\"otte$^{1}$ and Mark R Dennis$^2$}

\address{$^1$Max-Planck-Institute for the Physics of Complex Systems, Noethnitzer Str. 38, 01187 Dresden, Germany\\
$^2$H H Wills Physics Laboratory, University of Bristol, Tyndall Avenue, Bristol BS8 1TL, UK}

\begin{abstract} The magnitudes of beam shifts (Goos-H\"anchen and Imbert-Fedorov, spatial and angular) are greatly enhanced when a reflected light beam is postselected by an analyzer, by analogy with superweak measurements in quantum theory.
Particularly strong enhancements can be expected close to angles at which no light is transmitted for a fixed initial and final polarizations.
We derive a formula for the angular and spatial shifts at such angles (which includes the Brewster angle), and we show that their maximum size is limited by higher-order terms from the reflection coefficients occurring in the Artmann shift formula.
\end{abstract}



] 

\noindent  

The law of specular reflection fails for optical beams, as any physical beam is made up of multiple plane wave components, each experiencing a different reflection coefficient.
This variation leads to small deviations from perfect specular reflection: an angular shift in the beam's propagation direction \cite{Merano+:NatPhot3:2009}, and a spatial shift of the center of the reflected beam.
These beam shifts have attracted much experimental and theoretical interest in recent years (reviewed in \cite{BliokhAiello:JO14:2012}), consisting of shifts in the incidence plane (Goos-H\"anchen shift), perpendicular to it (Imbert-Fedorov shift), and of various modifications for different beam profiles (whilst assuming the incident intensity is axisymmetric).
In appropriate units, the shifts are independent of the particular radial beam profile, provided the beam is paraxial. 

As a wave optical phenomenon, beam shifts are typically small (a few wavelengths for spatial shifts, and approximately the beam's Fourier width for angular shifts). 
They are largest for incidence close to the critical angle or the Brewster angle, due to the special form of the Fresnel coefficients at these angles \cite{ChanTamir:OL10:1985,HorowitzTamir:JOSA:1971}.
Here we describe another way of enhancing the magnitude of beam shifts, namely by `postselecting' a particular polarization state with an appropriate choice of the polarization analyzer.
The payoff for the large apparent shift in the chosen polarization output is a small integrated intensity in that component; this consequence is standard in the quantum mechanical theory of `weak measurements' \cite{Aharonov+:PRL60:1988,Duck+:PRD40:1989,HostenKwiat:SCI:2008}, to which beam shifts are a classical analog \cite{GoetteDennis:NJP14:2012,DennisGoette:NJP14:2012}.

In this spirit, we define a \emph{null-reflection angle} with respect to a chosen initial and final polarization state, to be an angle of incidence for a reflected plane wave to be zero after the analyzer.
Such angles are somewhat analogous to the Brewster angle $\theta_{\mathrm{B}},$ at which plane waves polarized in the plane of incidence ($p$-polarized) are fully transmitted: the corresponding reflection coefficient $r_\mathrm{p}(\theta_{\mathrm{B}}) = 0,$ giving zero reflected light projected onto an analyzer set in the $p$-polarization state \cite{BornWolf:CUP:2003}.
Null-reflection angles are not polarizing angles \cite{Lakhtakia:ON15:1989,AzzamThonn:AO22:1983}, but they are interesting in the context of beam shifts, as they occur for arbitrary states of elliptic polarization, when the analyzer is crossed with (orthogonal to) the reflected polarization, and they can be found in regimes of both total and partial reflection. 

\emph{Beams}, made up of many plane wave components, do not have a homogenous polarization after reflection, and hence the projection onto an analyzer polarization is nonzero but possibly shifted.
With appropriate choice of initial and final polarizations, the beam shift close to Brewster or null-reflection angles can be very large. 
In the quantum analogy, this is the `superweak' regime \cite{BerryShukla:JPA43:2010}, where the value of the beam shift in the postselected component takes on a larger value than the shift of total intensity for either s or p incident polarization, albeit with a greatly reduced intensity.
For general incident and analyzer polarizations, the net component shift is arbitrary (it has a spatial and angular shift both with longitudinal and transverse components), and the apparent shift of the total beam is an appropriately weighted sum of orthogonal analyzer polarizations \cite{GoetteDennis:NJP14:2012}.

The magnitudes of optical beam shifts are usually calculated using the approximate Artmann formula \cite{Artmann:AndP437:1948} suitably generalized to angular, transverse and post-selected shifts \cite{GoetteDennis:NJP14:2012}.
At Brewster and null-reflection angles, this reflection coefficient is zero, meaning the Artmann formula blows up, implying that the shift is arbitrarily large.
A more refined calculation, given below, includes the next order in the Taylor expansion of the reflection coefficient, regularizing the singularity to be a resonance-like curve \cite{Kofman+:PhysRep520:2012,Gorodetski+:PRL109:2012} whose maxima and minima give the largest values of the beam 
shift, which now depend on the shape of the beam.
This behaviour is not unique to the Brewster angle, and our results extend previous theoretical work on Brewster reflection \cite{Aiello+:OL34:2009a,AielloWoerdman:arxiv:2007}.

For plane waves initial polarization and post-selecting analyzer are represented by Jones vectors $\boldsymbol{E}$ and $\boldsymbol{F}$ respectively, and the angle of incidence is $\theta.$
The reflection is described by a $2\times 2$ Jones matrix $\mathbf{R},$ which, in the p,s basis is given by $\mathbf{R} \equiv \left(\begin{smallmatrix} -r_\mathrm{p} & 0 \\ 0 & r_\mathrm{s} \end{smallmatrix}\right),$ where $r_\mathrm{p} = r_\mathrm{p}(\theta)$ $r_\mathrm{s} = r_\mathrm{s}(\theta)$ are the Fresnel reflection coefficients \cite{BornWolf:CUP:2003}.
A reflected plane wave will therefore have a polarization $\mathbf{R}\cdot\boldsymbol{E},$ and the amplitude after the analyzer is $\boldsymbol{F}^*\cdot\mathbf{R}\cdot\boldsymbol{E}.$
At $\theta_{\mathrm{B}},$ this is zero since $\boldsymbol{E}$ is $\boldsymbol{e}_\mathrm{p}$ and $r_\mathrm{p}(\theta_{\mathrm{B}}) = 0,$ giving zero transmission regardless of $\boldsymbol{F}.$
At null-reflection angles, $\boldsymbol{F}$ is orthogonal to the reflected polarization $\mathbf{R}\cdot\boldsymbol{E},$ and so any incident $\theta$ can be made into such an angle with an appropriate choice of $\boldsymbol{F}$ given $\boldsymbol{E}.$

A reflected beam is represented by a \emph{bundle} of plane waves, labelled by transverse wavevectors $\boldsymbol{K}$, centred on a central plane wave with wavevector $\boldsymbol{K} = \boldsymbol{0}$ in our chosen beam coordinate system.
Following the approach of Ref.~\cite{GoetteDennis:NJP14:2012}, it is convenient to work in cylindrical coordinates with axis the 'virtual reflected beam' (i.e.~unshifted), with azimuth $\phi$ and radius $r$ in real space, azimuth $\alpha$ and radius $\sin \delta$ in Fourier space. 
The shift formulas for a reflected beam follow from a Taylor expansion of 
\begin{equation}
   \rho(\boldsymbol{K}) = \boldsymbol{F}^*\cdot\mathbf{R}(\boldsymbol{K})\cdot\boldsymbol{E},
   \label{eq:rhodef}
\end{equation}
where $\mathbf{R}(\boldsymbol{K})$ is the $\boldsymbol{K}$-dependent reflection matrix, which is a 2 $\times$ 2 Jones matrix in beam coordinates, and $\mathbf{R}(0)$ is the reflection matrix of the central plane wave.
Since the beam is paraxial, its spectrum is localized around the origin in Fourier space, justifying the expansion 
of $\rho(\boldsymbol{K})$ up to only the first few terms \cite{BliokhAiello:JO14:2012},
\begin{equation}
   \rho(\boldsymbol{K}) \approx \rho_0 + \delta \boldsymbol{\rho}_1 \cdot \binom{c(\alpha)}{s(\alpha)} + \tfrac{1}{2} \delta^2 \binom{c(\alpha)}{s(\alpha)}\cdot \mathbf{P}_2 \cdot \binom{c(\alpha)}{s(\alpha)},
   \label{eq:rhoexp}
\end{equation}
where $c(\alpha) = \cos \alpha, s(\alpha) = \sin \alpha,$ and $\rho_0, \boldsymbol{\rho}_1$ and $\mathbf{P}_2$ all originate from the Taylor expansion of $\rho$.
We will implicitly use the $\boldsymbol{e}_\mathrm{p}, \boldsymbol{e}_\mathrm{s}$ basis, with polar coordinates $r, \phi$ in real space and $K = K(k \delta, \alpha)$ in Fourier space (with the range of $\delta$ so small that $\sin \delta \approx \delta$).
The unperturbed, virtual amplitude (which corresponds to the incident beam) is real and axisymmetric, and is written $\varphi(r),$ with Fourier transform $\widetilde{\varphi}(K).$
We assume that these are normalized, $\int_0^{\infty} r |\varphi(r)|^2 dr = \int_0^{\infty} K |\widetilde{\varphi}(K)|^2 dK = 1.$

\begin{figure}
  \begin{center}
   \includegraphics[width=\linewidth]{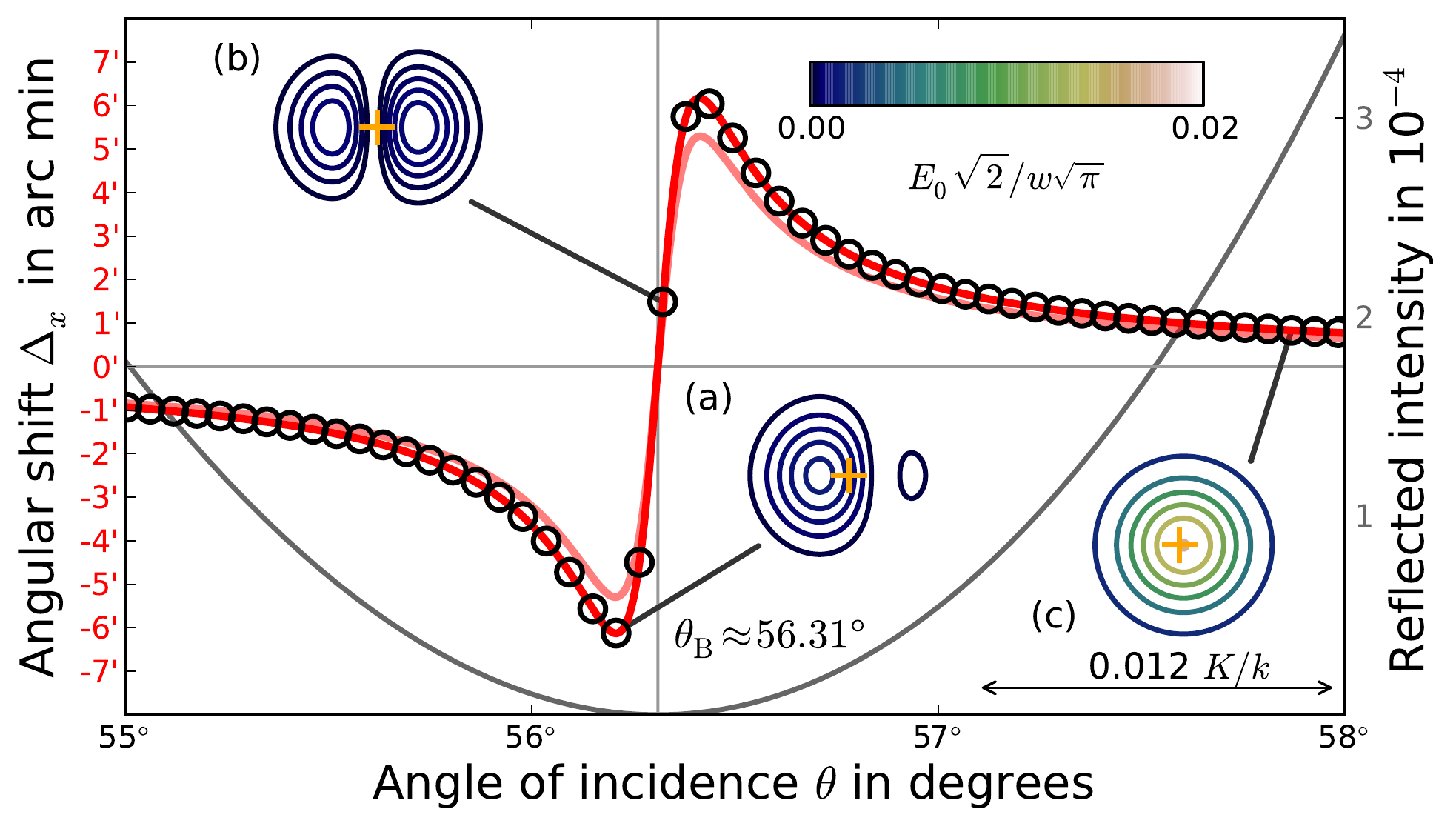}
   \caption{ 
   Angular shifts close to a conventional Brewster angle.
   The $\theta$-dependent angular shift $\Delta_x$ (thick, red curve), with $\boldsymbol{F} = (1,0)$ is compared with the shift of the total intensity \cite{Merano+:NatPhot3:2009} (light pink curve) for a reduced refractive index of $n=1.5$ (e.g. glass-air interface).
   Both shifts follow a resonance-like curve close to the Brewster angle at $\theta_\mathrm{B} = 56.31^\circ$. 
   Black circles indicate $\Delta_x$ obtained by numerical integration of (\ref{eq:Deltadef}). 
   The moments of the spectrum $\langle \delta^2 \rangle$ and $\langle  \delta^4 \rangle$ are obtained for a Gaussian beam with beam waist of 50 $\mu$m and a wavelength of 0.526 $\mu$m.
   On the same abscissa, we plot the ratio of reflected to incident intensity (thin, gray curve), showing a minimum close to the maximum shift. 
   The figure also shows three contour plots (insets) of the angular spectrum of the light beam at angles $\theta = 56.02^\circ$(a), $56.32^\circ$(b) and $57.87^\circ$(c). 
   Orange crosses indicate the origin from which the shifts are measured. 
   The strength of the contour is shown in the colorbar in the top right corner.}
   \label{fig:angular}
 \end{center}  
\end{figure}

The beam shift is determined by the centroid (position expectation value) of the reflected beam $\psi(\boldsymbol{r})$, which is no longer axisymmetric.
In Fourier space, $\rho(\boldsymbol{K})$ is a multiplication operator on the amplitude, and $\widetilde{\psi}(\boldsymbol{K}) = \rho(\boldsymbol{K})\widetilde{\varphi}(K).$
A good approximation of the angular shift $\Delta,$ valid for a paraxial beam close to the Brewster and null-reflection angle, comes from 
applying the expansion (\ref{eq:rhoexp}) to this expectation, i.e.~
\begin{eqnarray}
   \Delta & = & \frac{1}{k}\frac{\int_0^\infty dK \, K \int_0^{2\pi}d \alpha\, \boldsymbol{K} |\widetilde{\psi}(\boldsymbol{K})|^2}{\int_0^\infty dK\, K \int_0^{2\pi}d \alpha\, |\widetilde{\psi}(\boldsymbol{K})|^2} 
   \label{eq:Deltadef} \\
   & \approx  &\frac{\int_0^\infty d\delta \int_0^{2\pi}d \alpha\, \delta^2 (\cos\alpha,\sin\alpha) |\widetilde{\psi}(\boldsymbol{K})|^2}{\int_0^\infty d\delta \int_0^{2\pi}d \alpha\, |\widetilde{\psi}(\boldsymbol{K})|^2} \nonumber \\
   & \approx  &\langle \delta^2\rangle \frac{\mathrm{Re}\left\{\rho_0^* \boldsymbol{\rho}_1^{\vphantom{*}} + \tfrac{\langle\delta^4\rangle}{8\langle\delta^2\rangle}(2 \mathbf{P}_2^{\vphantom{*}} \cdot \boldsymbol{\rho}_1^*+ \boldsymbol{\rho}_1^* \mathrm{tr}\mathbf{P}_2^{\vphantom{*}} \right\}}{|\rho_0|^2+\langle\delta^2\rangle \left[ \tfrac{ \| \boldsymbol{\rho}_1^{\vphantom{*}}\|} {2} + \mathrm{Re} \left( \rho_0^* \mathrm{tr} \mathbf{P}_2^{\vphantom{*}} \right) \right]}.
   \label{eq:Delta}
\end{eqnarray}
In this expression we have omitted small terms such as those combining $\rho_0$ with third order derivatives of $\rho$.
The usual Artmann formula for this case contains only terms with $\rho_0$ and $\boldsymbol{\rho}_1$.
In the last line, $\langle \delta^n \rangle \equiv k^3 \int_0^{\infty} d\delta \, \delta^{n+1}|\widetilde{\varphi}(\delta)|^2,$ the $n$th 
radial moment of the spectrum, and $\langle \delta^{n+1} \rangle \ll \langle \delta^{n} \rangle$ for each $n,$ normalized such that $\langle \delta^0 \rangle = 1.$
Thus, in usual beam shift circumstances, only the first term in the numerator and denominator contributes to the angular shift, giving the known angular shift formula \cite{Merano+:NatPhot3:2009}, which is independent of the beam profile apart from the overall scaling by the second moment of the spectrum.
However, when $|\rho_0| \ll 1,$ close to the Brewster and null-reflection angle, the higher terms included in Eq.~(\ref{eq:Delta}) 
contribute, replacing the singularity with a resonance-like complex singularity, as shown in Fig.~\ref{fig:angular}.
The postselection enhances the already large Brewster angle shift for total intensity, which follows a similar curve \cite{AielloWoerdman:arxiv:2007}.
Beam shifts at null-reflection angles can be both angular and spatial, and in an arbitrary direction in the $xy$-plane 
\cite{GoetteDennis:NJP14:2012}; the shift is purely angular, in the plane of incidence at real Brewster angle.

The calculation of the formula for the spatial shift $\boldsymbol{D}$ at the null-reflection angle is similar to the angular calculation 
above, but with $\varphi(r)$ replacing its Fourier transform.
However, in real space, the reflection operator $\rho$ becomes a differential operator with $\boldsymbol{K} \to -i \boldsymbol{\nabla}$ in the expansion (\ref{eq:rhoexp}), giving $\psi(\boldsymbol{r}) = \rho(-i \boldsymbol{\nabla})\varphi(r).$
The action of the operator in Taylor expanded form on $\varphi(r)$ is easy to calculate, and so the formula for the regularized spatial shift $\boldsymbol{D}$ is
\begin{eqnarray}
   \boldsymbol{D} & = & \frac{\int_0^\infty dr \, r \int_0^{2\pi}d \phi\, \boldsymbol{r} |\psi(\boldsymbol{r})|^2}{\int_0^\infty dr\, r \int_0^{2\pi}d \phi\, |\psi(\boldsymbol{r})|^2} 
   \label{eq:Ddef} \\
   & \approx & -\frac{1}{k} \frac{\mathrm{Im}\left\{\rho_0^* \boldsymbol{\rho}_1^{\vphantom{*}} +\tfrac{\langle \delta^2 \rangle}{4}\left( 2\mathbf{P}_2^{\vphantom{*}} \cdot\boldsymbol{\rho}_1^*+ \boldsymbol{\rho}_1^* \mathrm{tr}\mathbf{P}_2^{\vphantom{2}} \right)\right\}}{|\rho_0|^2+ \langle \delta^2 \rangle \left[ \tfrac{ \| \boldsymbol{\rho}_1^{\vphantom{*}}\|} {2} + \mathrm{Re} \left( \rho_0^* \mathrm{tr} \mathbf{P}_2^{\vphantom{*}} \right) \right]},
   \label{eq:D}
\end{eqnarray}
omitting the technical but straightforward integrals and truncating at second order derivatives of $\rho$.
Unlike the usual beam shift, the regularized spatial shift now depends on the beam's Fourier width $\langle \delta^2 \rangle$.
As before, when $\rho$ is not small, $\boldsymbol{D}$ is given by Artmann formula for the spatial shift \cite{Artmann:AndP437:1948}.
However, when $|\rho_0| \ll 1,$ once again the next terms in numerator and denominator play a role, regularizing the infinite expression in the usual Artmann formula.

\begin{figure}
  \begin{center}
   \includegraphics[width=\linewidth]{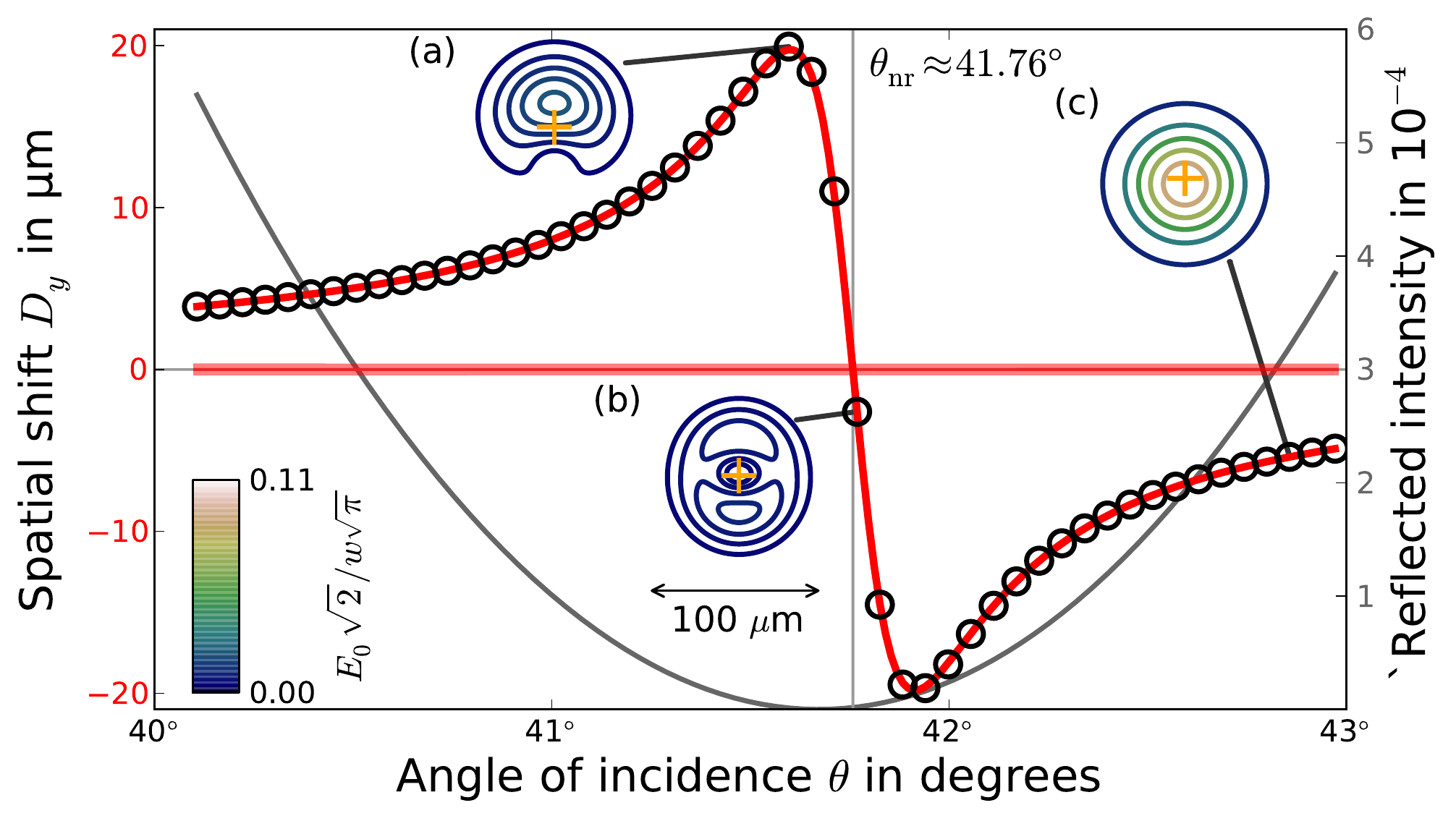}
   \caption{\label{fig:spatial}
   Large, weak spatial shift close to a null-reflection angle $\theta_\mathrm{nr}$.
   The shift $D_y$ (thick, red curve) is a function of $\theta$ for a refractive index of $n=3/10,$ showing the resonance-like curve around $\theta_\mathrm{nr} \approx 41.76^\circ$.
   The light pink curve is the shift for Imbert-Fedorov eigenpolarization which is smaller than $0.15\:\mu$m across the shown range.
   On the same axis, the ratio of reflected to incident intensity is plotted (thin, gray curve). 
   The black circles are the shift for a numerical experiment for a Gaussian beam with a waist of 50 $\mu$m and a wavelength of 0.526 $\mu$m.
   Inset are contour plots of the beam's amplitude at angles $\theta = 41.60^\circ$(a), $41.77^\circ$(b) and  $42.86^\circ$(c). 
   Orange crosses indicate the origin for the contour plots. 
   }
 \end{center}  
\end{figure}

To illustrate the amplification in the superweak regime, in Fig.~\ref{fig:spatial} we plot the post-selected Imbert-Fedorov shift around a null-reflection angle for initial circular and final 45$^{\circ}$ linear polarizations, which again follows a resonance-like curve.
This shift is much larger than the largest possible Imbert-Fedorov shift in the total intensity at this angle, which occurs for eigenpolarizations of the Imbert-Fedorov effect \cite{Fedorov:DAN105:1955}.
Because the weak, postselected shift is larger than this shift, we truly are in the \emph{superweak} regime, where weak values are far larger than the spectrum of the operator. 
The concurrent low intensity, however, is still experimentally accessible \cite{Jayaswal+:OL38:2013}.
We note that this choice of polarizations leads to an optical vortex in the reflected beam.

For usual beam shifts, the spatial and angular shifts are the real and imaginary part of the same complex expression, which may be interpreted as a complex weak value of the Artmann operator \cite{DennisGoette:NJP14:2012,Josza:PRA76:2007,Kobayashi+:PRA86:2012}.
However, this is no longer the case for a general beam profile, as the weighting on the spectral moments is different in Eqs.~(\ref{eq:Delta}) and (\ref{eq:D}).
This discussion illustrates another fact about shifts close to the Brewster angle: in going beyond first order, the detailed behavior of the shift becomes dependent on the spectral profile, unlike the regular beam shift.
We note that this is not necessarily the case for higher-order beam shift phenomena, as the case of a high-order optical vortex on reflection demonstrates \cite{DennisGoette:PRL109:2012}.

\pagebreak

\section*{Informational Fourth Page}
In this section, please provide full versions of citations to
assist reviewers and editors (OL publishes a short form of
citations) or any other information that would aid the peer-review
process.

\end{document}